\documentclass[
               biblatex,     
               ]{jacow}

\usepackage[english]{babel}

\begin{document}

\title{Using Chromaticity to Suppress Instabilities at $\gamma_t$ Transition}

\author{M.~A.~Balcewicz\thanks{balcewic@fnal.gov}, J.~Eldred, Fermi National Accelerator Laboratory, Batavia, United States}

\maketitle


\begin{abstract}
For many synchrotrons which cross $\gamma_t$ transition, a high intensity threshold has been identified beyond which significant beam losses are observed. These losses are due to a high-intensity mode coupling instability, which can be suppressed by adjusting the chromaticity curve to cross zero slightly after $\gamma_t$ transition.
\end{abstract}

\section{Introduction}



The Fermilab Booster crosses the transition energy $\gamma_t$ approximately halfway through the energy ramp. For sufficient bunch intensity when crossing transition large dipole oscillations and losses are observed in the bunch tail. This loss is believed to be due to a beam instability which is excited when synchrotron oscillations slow around $\gamma_t$ transition. These dipole oscillations are concentrated around the tail of the bunch. Below the loss threshold tail dominated dipole oscillations can be observed using a dipole kicker.

The intensity threshold for losses is above 6.0$\times 10^{12}$ ppp (particles per pulse, each pulse has 81 bunches). PIP-II and LBNF/DUNE upgrades require 6.7$\times 10^{12}$ ppp with $\geq95\%$ of the beam extracted, making it necessary to mitigate this beam instability.

\section{Transition Instability}

The observed high-intensity instability is believed to be a form of Transverse Mode Coupling Instability (TMCI), with the large tail amplitude created by wakes accumulating along the bunch length. This amplitude is partially stabilized by space charge forces. The large tail amplitude is variously understood to be a characteristic of TMCI at very strong space charge\cite{zermat}, the Convective Instability\cite{burovABS}, or Beam Break Up\cite{cernps} occurring much more rapidly than the synchrotron oscillations.

To reach the required intensity it is necessary to minimize growth due to TMCI. At lower intensities, space charge can decouple TMCI modes and stabilize the beam\cite{blaskiewicz1}. However, in the limit where $\Delta Q_{sc}/Q_s\to\infty$, space charge instead makes the instability threshold vanishingly small\cite{zermat}. Near transition where the synchrotron tune is small, space charge cannot be used to minimize emittance growth. A  small nonzero chromaticity can be used to dampen TMCI. However, the beam is unstable to Head-Tail when the effective chromaticity $\xi_{eff}\equiv\xi/\eta<0$, where $\xi$ is the chromaticity and $\eta$ is the phase slip factor. Without a $\gamma_t$ jump system the beam will at least be unstable to TMCI or Head-Tail near transition.

Instabilities do not need to be fully stabilized to avoid particle losses. Instead it is sufficient limit the dipole amplitude such that minimal beam is lost to the aperture. With this objective in mind, the total growth of coherent modes and their respective tail amplification can be optimized and a new working point can be determined for high-intensity operations.

In order to find a new working point, a simulation campaign has been undertaken to better understand the functional dependence of transition crossing instabilities in the Fermilab Booster.

\section{Semi-Analytic Instability Model}

The semi-analytic Multiple Loop Square Well (MLSW) model\cite{thesis} is used to simulate instabilities over possible transition crossing schemes and determine the coherent modes and growth rates. This model considers linear dynamics with space charge, wakes, synchrotron tune, and chromaticity. Longitudinal coupling is provided by a layered longitudinal phase space within nested square potential wells. 

With fully coherent transverse motion, the system determines the tune shift of coherent synchro-betatron modes and their respective dipole amplitudes (the eigenvalues and eigenvectors). By stitching together the solutions at settings across transition it is possible to approximate how coherent modes evolve, but not calculate the relative contribution of each mode.
\begin{figure}[!htb]
   \centering
   \includegraphics*[width=3.0in]{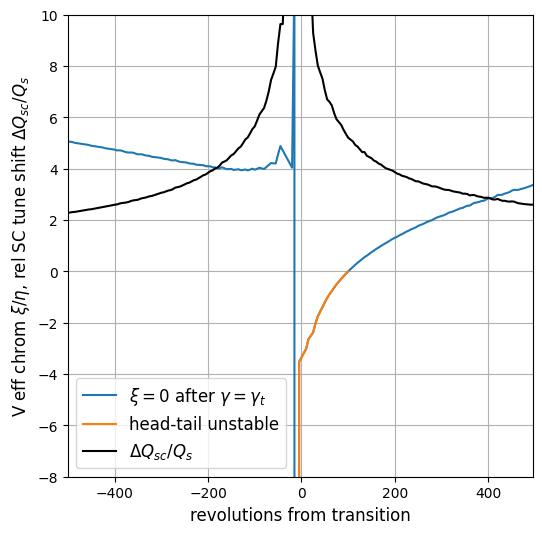}
   \caption{Vertical chromaticity and synchrotron tunes near beam transition. Other parameters can be considered approximately constant within this region. For these settings the vertical chromaticity curve crosses 100 revolutions after transition driving Head-Tail instability.}
   \label{fig:simulated curves}
\end{figure}

Instabilities of this type occur within a 1000 revolution ($\sim$1.6 ms) region centered on beam transition, so simulations are constrained to this region of parameter space. Vertical chromaticity is not well measured at transition, but is assumed to be linear with a slope inferred from measurements taken on either end of transition and can be offset to model Head-Tail. The synchrotron tune over the ramp can found using the Wall Current Monitor (WCM) system. The resulting effective chromaticity and relative synchrotron tune shift ($\Delta Q_{sc}/Q_s$) are shown in fig. \ref{fig:simulated curves}.
\begin{figure}[!htb]
   \centering
   \includegraphics*[width=3.0in]{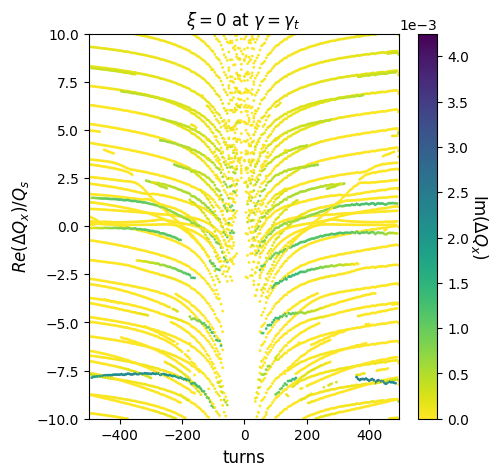}
   \caption{Simulated TMCI and growth rates for individual coherent modes assuming vertical chromaticity curve crosses zero at transition}
   \label{fig:nom tmci}
\end{figure}

Simulations were performed at different beam intensities and chromaticity offsets. Intensity has three main effects on beam instability, increasing the unstable region and corresponding growth rate of unstable modes. At the same time, the head tail amplitude also increases with beam intensity. Fig. \ref{fig:nom tmci} shows that when $\Delta Q_{sc}/Q_s$ is sufficiently large, space charge does not stabilize TMCI\cite{zermat}. This effectively means that there is no well-defined intensity threshold for beam instability; however, there is a threshold below which growth is heavily suppressed and losses do not occur.
\begin{figure}[!htb]
   \centering
   \includegraphics*[width=3.0in]{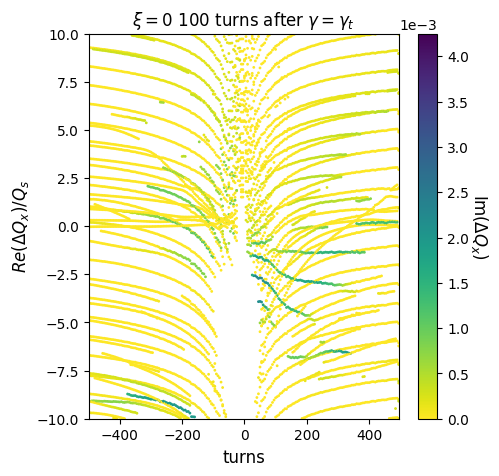}
   \caption{Simulated TMCI and Head-Tail growth rates for individual coherent modes using the chromaticity and synchrotron curves in fig. \ref{fig:simulated curves}.}
   \label{fig:growth rate}
\end{figure}

\begin{figure}[!htb]
   \centering
   \includegraphics*[width=3.0in]{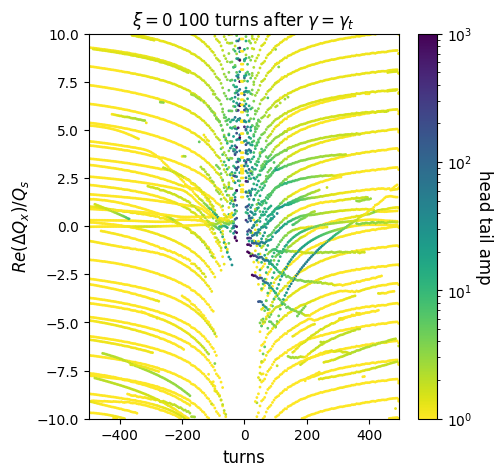}
   \caption{Simulated at the same settings Ratio of maximum tail amplitude to initial head amplitude for individual coherent modes using the chromaticity and synchrotron curves in fig. \ref{fig:simulated curves}. Model assumes steady state; actual beam will not achieve maximum head tail amplitude. Amplification is largest at transition when synchrotron oscillation is smallest, even with a nonzero chromaticity at transition.}
   \label{fig:head tail amp}
\end{figure}

Chromaticity on the other hand is shown to dampen TMCI growth around beam transition, but creates a weaker but longer lived Head-Tail instability either before or after beam transition. This is shown in fig. \ref{fig:growth rate}. While instability growth accumulates over time, the head tail amplification is only large near beam transition even when chromaticity is shifted as in fig. \ref{fig:head tail amp}.

This suggests that it isn't actually necessary to fully stabilize the beam at transition. Instead, beam instabilities can be delayed past the maximum head tail amplification found at transition. Doing this will significantly decrease the dipole amplitude of these instabilities and diminish particle losses.




\section{Experimental Observations}
If results of the simulations are valid, it should be possible to suppress losses across high-intensity ($\geq6\times 10^{12}$ ppp) beam transition by shifting the vertical chromaticity curve to cross zero slightly after beam transition. The maximum intensity achievable by the Fermilab Booster is set by current from the normal conducting linac. This varies over time, but can be configured to provide higher intensity in exchange for degraded beam quality. This can only be done in dedicated studies as it requires significant linac tuning and will increase losses in Booster.

\begin{figure}[!htb]
   \centering
   \includegraphics*[width=3.0in]{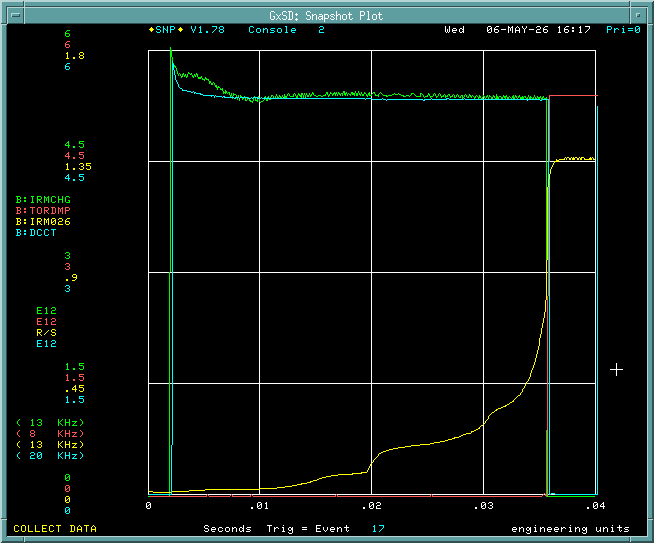}
   \caption{Maximum beam intensity of 5.3$\times 10^{12}$ ppp achievable at nominal settings.}
   \label{fig:experimenthi}
\end{figure}

At current operational conditions the maximum beam intensity achievable was 5.3$\times 10^{12}$ ppp. at transition. As with prior studies, this was not sufficient to excite a dipole instability (fig. \ref{fig:experimenthi}) until vertical chromaticity was offset by over 1 ms in either direction and a strong Head-Tail instability was observed.

\begin{figure}[!htb]
   \centering
   \includegraphics*[width=3.0in]{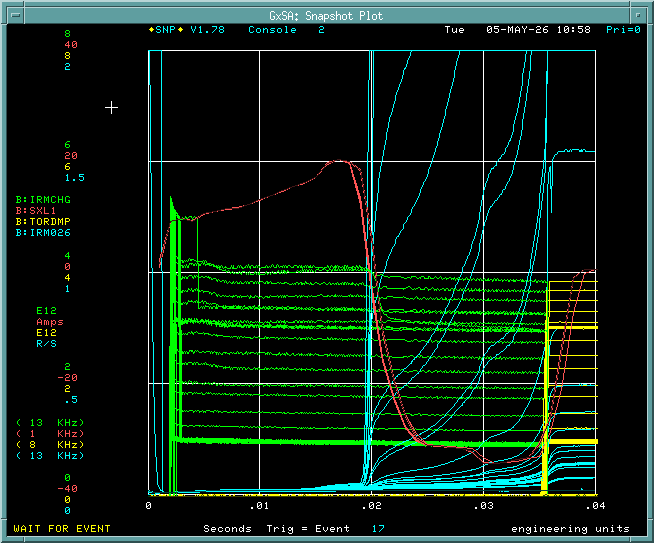}
   \caption{Beam intensity along the ramp for a periodically kicked beam. Losses eventually observed at 16 turns of intensity (4$\times 10^{12}$ ppp). Although this kick is sufficient to drive losses, it does not seem to fully excite beam instability.}
   \label{fig:experiment}
\end{figure}

Prior studies have shown that it is possible to excite large dipole oscillations in the tail of the bunch with a dipole kicker. This is normally used to excite oscillations around the closed orbit to measure beam tune including directly before transition. When excited this way, significant losses are observed, and beam intensity should be kept to the minimum intensity needed to drives these losses. With this kicker beam loss is consistent with an instability was observed at 4$\times 10^{12}$ ppp shown in fig \ref{fig:experiment}. Although this loss could be excited, optimization of chromaticity did not change the loss profile. Since intensity makes such a major contribution to growth rates and head tail amplification, these results suggest the chromaticity mitigation strategy may become relevant only at $\sim6\times 10^{12}$ ppp where instabilities do not need to be excited by a dipole kicker.

\section{Conclusion}

Semi-analytic modeling of the Fermilab Booster suggests that high-intensity mode coupling instabilities around $\gamma_t$ transition can be minimized by shifting the vertical chromaticity curve to cross zero slightly after transition. However, recent studies conducted at the Fermilab Booster were unable to achieve the needed beam intensity to excite the instability in question. Dedicated experiments at higher intensity are required to confirm that transition instabilities can be damped in this way and whether other mitgations like active damping or a $\gamma_t$ jump system are needed to reach $6.7\times10^{12}$ ppp.

In conjunction with experimental work, the simulations will be further improved. The current wake strength is believed to be approximately correct, but has a functional form defined by laminations within the main combined function magnets\cite{laminations}. In addition, a newly developed lattice model promises to improve our understanding of how chromaticity and other optics vary near transition.

\printbibliography

\end{document}